\documentclass[reprint,superscriptaddress,amsmath,amssymb,prb,longbibliography]{revtex4-1}
\usepackage{comment}
\usepackage{color}
\usepackage{graphicx}
\usepackage{subfigure}
\usepackage{dcolumn}
\usepackage{bm}
\usepackage{amsmath}
 
\usepackage{braket}
\usepackage{diagbox, eqparbox, hhline}
\usepackage[detect-none]{siunitx}
\DeclareMathAlphabet{\mathpzc}{OT1}{pzc}{m}{it}

\newcommand{\Keff}{K_{\text{eff}}} 
\newcommand{\Msat}{M_{\text{s}}}

\begin{document}


\title{Manipulation of Magnetic Skyrmions by Superconducting Vortices in Ferromagnet-Superconductor Heterostructures }
\author{Ra\'i M. Menezes}
\affiliation{%
Departement Fysica, Universiteit Antwerpen, Groenenborgerlaan 171, B-2020 Antwerpen, Belgium
}
\affiliation{%
Departamento de F\'isica, Universidade Federal de Pernambuco, Cidade Universit\'aria, 50670-901, Recife-PE, Brazil
}%
\author{Jos\'e F. S. Neto}
\affiliation{%
Departamento de F\'isica, Universidade Federal de Pernambuco, Cidade Universit\'aria, 50670-901, Recife-PE, Brazil
}%
\author{Cl\'ecio C. de Souza Silva}
\affiliation{%
Departamento de F\'isica, Universidade Federal de Pernambuco, Cidade Universit\'aria, 50670-901, Recife-PE, Brazil
}%
\author{Milorad V. Milo\v{s}evi\'c}%
\email{milorad.milosevic@uantwerpen.be}
\affiliation{%
Departement Fysica, Universiteit Antwerpen, Groenenborgerlaan 171, B-2020 Antwerpen, Belgium
}%

\date{\today}

\begin{abstract}
Dynamics of magnetic skyrmions in hybrid ferromagnetic films harbors novel physical phenomena and holds promise for technological applications. In this work, we discuss the behavior of magnetic skyrmions when coupled to superconducting vortices in a ferromagnet-superconductor heterostructure. We use numerical simulations and analytic arguments to reveal broader possibilities for manipulating the skyrmion-vortex dynamic correlations in the hybrid system, that are not possible in its separated constituents. We explore the thresholds of particular dynamic phases, and quantify the phase diagram as a function of the relevant material parameters, applied current and induced magnetic torques. Finally, we demonstrate the broad and precise tunability of the skyrmion Hall-angle in presence of vortices, with respect to currents applied to either or both the superconductor and the ferromagnet within the heterostructure.
\end{abstract}

\pacs{Valid PACS appear here}
\maketitle

\section{Introduction}

The ability to trap and manipulate magnetic skyrmions is of great recent importance for cutting-edge memory devices and information technology\cite{ParkinMagneticMemory,Fert2013SkyrmionsTrack,Kiselev2011ChiralTechnologies,Nagaosa2013TopologicalSkyrmions}. Magnetic skyrmions are topologically protected spin textures which can be stabilized, e.g, in ultrathin ferromagnetic films when coupled to a heavy metal (HM) layer with strong spin-orbit coupling. The broken interfacial inversion symmetry induced by the heavy-metal layer produces an interfacial non-collinear Dzyaloshinskii-Moriya interaction (DMI), which energetically favors N\'eel-type skyrmions and domain walls\cite{jiang2017skyrmions, fert2017magnetic,Bogdanov1994ThermodynamicallyCrystals,Bogdanov2001ChiralMultilayers}. 

\begin{figure}[t]
\centering
\includegraphics[width=\linewidth]{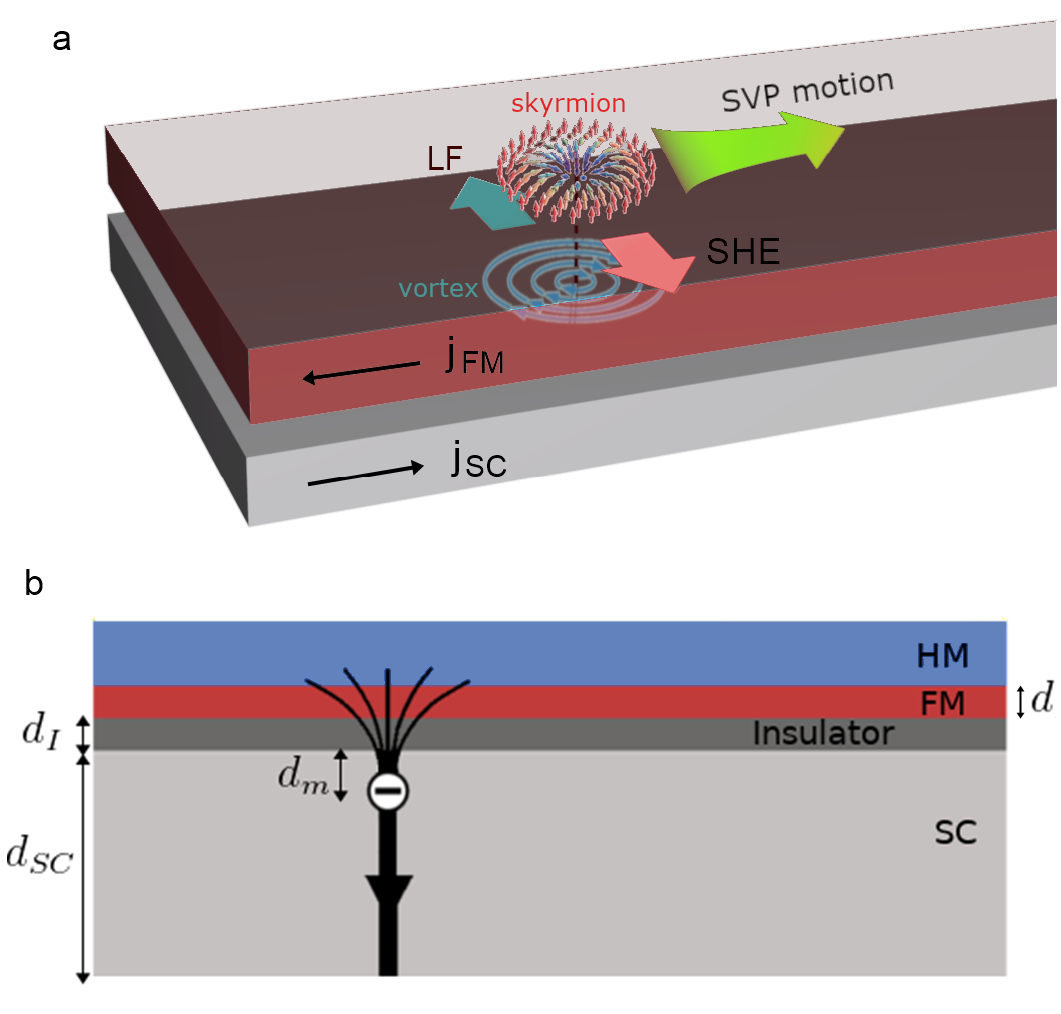}
\caption{(a) Oblique view of the considered heterostructure. By tuning the competition between the Lorentz force (LF), acting on the superconducting vortex, and the magnetic torques acting on the skyrmion, one can control the resultant skyrmion Hall effect (SHE) and the net direction of the skyrmion-vortex pair (SVP) motion. (b) Schematic details of the considered system, a thin ferromagnetic (FM) film of thickness $d$ with perpendicular magnetic anisotropy, coupled to a heavy metal (HM) layer with a strong spin-orbit coupling, placed on top of a superconducting (SC) film of thickness $d_{\text{SC}}$, separated by an insulating layer of thickness $d_\text{I}$, such that the interaction between the superconducting and the ferromagnetic film is restricted to only the magnetic stray fields.}
\label{fig1}
\end{figure}

Heterostructures often present nontrivial phenomena enabled by the competition or hybridization of the physical properties of its parts. Particularly, ferromagnet-superconductor (FM-SC) heterostructures have received much attention in recent years\cite{blamire2014interface,bergeret2005odd,eschrig2015spin,buzdin2005proximity}, either for their possible applications in spintronics\cite{linder2015superconducting} and Josephson devices\cite{halasz2009critical,kalenkov2011triplet,yokoyama2015josephson,alidoust2015proximity}, or for the rich emergent physics in such systems\cite{zorro2014nucleation,pershoguba2016skyrmion,bjornson2014skyrmion,helseth2002interaction,vadimov2018magnetic}. Recently, theoretical works on chiral FM-SC heterostructures have demonstrated that the stray magnetic field of superconducting vortices may be able to create\cite{del2015imprinting} magnetic skyrmions in the ferromagnetic layer, also to trap or repel the preexisting skyrmions \cite{hals2016composite,dahir2019interaction}, depending on vortex polarity. First insights in the dynamic properties of such hybrid systems were recently provided in Ref. \onlinecite{hals2016composite}.  Here, we provide an in-depth analysis and investigate the manipulation of the skyrmion-vortex pair (SVP) correlations in a FM-SC hybrid, in case of independently biased films (current applied to either FM or SC part). We study the dependence of the net motion of skyrmions and vortices on the viscosities of the host materials, the exerted Lorentz force and magnetic torques by applied current(s), and calculate the skyrmion Hall-angle with respect to currents applied into both superconductor and ferromagnetic films. We reveal that the skyrmion Hall-angle with respect to current applied into the ferromagnetic film is always greater than one observed in the absence of vortices. We stress the possibility of compensating the skyrmion Hall effect (SHE) in such systems by applying combined currents into two constituent materials of the heterostructure, which is of importance for the facilitated skyrmion guidance in racetrack applications, where the SHE can cause skyrmion to annihilate at the sample edges. Fig.~\ref{fig1} illustrates the considered system, an ultrathin ferromagnetic film of thickness $d$ with perpendicular magnetic anisotropy, e.g., a Co layer, coupled to a nonmagnetic layer on top with a strong spin-orbit coupling, e.g., the heavy metal Pt (neither Co nor Pt are superconductors at ambient pressure), placed on top of a superconducting film of thickness $d_{\text{SC}}$, separated by an insulating layer of thickness $d_\text{I}$, such that the interaction between the superconducting material and the ferromagnetic film is solely through the magnetic stray fields.    

The paper is organized as follows. In Sec.~\ref{Sec.II} we provide analytic considerations before describing the micromagnetic model of ferromagnetic films with interfacially-induced DMI and providing the Thiele formalism for the center-of-mass motion of the magnetic skyrmion. In Sec.~\ref{Sec.III} we report the static properties of the hybrid system in presence of superconducting vortices, i.e. the general considerations of the ferromagnetic state in the stray field of a vortex, the properties of the skyrmion, and the skyrmion-vortex interaction. Sec. \ref{Sec.IIIb} is devoted to dynamic properties of the hybrid system, where we combine micromagnetic and molecular dynamics simulations to investigate the behavior of skyrmions and vortices simultaneously when currents are applied into both SC and FM part of the heterostructure. In Secs. \ref{S.IV.B}-\ref{sec.dynamics-CurrSC} we consider an uniform current applied only to the superconductor, where we show the dependence of the dynamic phases on the material viscosities and calculate the critical properties of the SVP, as well as the angle of the SVP terminal motion with respect to the applied current. In Sec.~\ref{S.IV.C} we show that the skyrmion Hall-angle with respect to currents applied into the FM film is always greater than that observed in the absence of vortices, and describe the full potential of guiding the magnetic skyrmions by tuning the skyrmion Hall effect in FM-SC hybrid systems. Our results are summarized in Sec.~\ref{Sec.IV}.
 
\section{Theoretical formalism}\label{Sec.II}
In this work, we rely on molecular dynamics simulations and the London limit to describe the vortex behavior in the superconducting layer. Then the stray magnetic field of the (moving) vortices is used in the micromagnetic framework to understand the static and dynamic response of the ferromagnetic layer and skyrmions therein. For description of the dynamic phases of the heterostructure as a whole, we couple the molecular dynamics of vortices with the Thiele equation of motion of skyrmions. In what follows, we give a short description of the key ingredients in our theoretical analysis.

\subsection{Stray field of a single vortex}

The stray field of the superconducting vortex can be calculated analytically in the London limit, $\lambda\gg\xi$, where $\lambda$ and $\xi$ are the penetration depth and the coherence length, respectively. The general solution for  the stray field produced outside the superconducting film of thickness $d_\text{SC}$ by a straight vortex reads\cite{carneiro2000vortex} 
\begin{subequations}\label{eq.1}
\begin{align}
    B_r(r,z>0)=&\frac{\phi_0}{2\pi\lambda^2}\int_0^{\infty}dk\frac{kJ_1(kr)}{k^2+\lambda^{-2}}f(k,z), \label{eq0a}\\
    B_z(r,z>0)=&\frac{\phi_0}{2\pi\lambda^2}\int_0^{\infty}dk\frac{kJ_0(kr)}{k^2+\lambda^{-2}}f(k,z), \label{eq0b}
\end{align}
\end{subequations}
where $$f(k,z)=\tau e^{-kz}\frac{(k+\tau)e^{\tau d_\text{SC}}+(k-\tau)e^{-\tau d_\text{SC}}-2k}{(k+\tau)^2e^{\tau d_\text{SC}}-(k-\tau)^2e^{-\tau d_\text{SC}}},$$
and $\tau=\sqrt{k^2+\lambda^{-2}}$. Here, $z=0$ represents the superconductor surface and $r=\sqrt{x^2+y^2}$ the distance from the center of the vortex core.
As discussed in Ref. \onlinecite{carneiro2000vortex}, for the case of $d_\text{SC}\gg\lambda$, the stray field of a single vortex can be approximated, near the superconductor surface, by the field of a magnetic monopole of ``charge'' $2\phi_0$, where $\phi_0$ is the magnetic flux quantum, located at a distance $d_\text{m}=1.27\lambda$ below the superconductor surface.  In this case, the stray field takes the simple form
\begin{subequations}\label{eq.2}
\begin{align}
    B_r(r,z>0)=&\frac{\phi_0}{2\pi}\frac{r}{R^3}, \label{fig1a}\\
    B_z(r,z>0)=&\frac{\phi_0}{2\pi}\frac{z+d_m}{R^3}, \label{fig1b}
\end{align}
\end{subequations}
where $R=\sqrt{r^2+(z+d_m)^2}$ is the distance from the monopole. We use this approximation in our calculations for the case of thick superconducting films, $d_{\rm SC}\gg\lambda$, while for small or moderate thicknesses we use the full expression given by Eq.~\ref{eq.1}. For more details, refer to Appendix~\ref{AppB}.

In our system, the ferromagnetic film is placed on top of the superconductor, separated by an insulating layer of thickness $d_I$, thus experiencing the stray field of the superconducting vortex calculated in the plane $z=d_{I}$. We consider an ultrathin FM film, such that the DMI, induced in the FM-HM interface, and the magnetic field induced by the vortex, are considered to be uniform across the film thickness. Notice that in this work we do not consider the creation of vortex-antivortex pairs in the superconductor due to the stray field of the skyrmion\cite{dahir2019interaction,baumard2018generation}, since such a stray field emanating from an ultrathin FM film is insufficient to strongly perturb the superconducting film separated by a thick insulating layer.

\subsection{Micromagnetic model}\label{Sec.IIb}

For the micromagnetic simulations of the chiral ferromagnetic layer, we employ the simulation package mumax$^3$ (see Refs. \onlinecite{Vansteenkiste2014TheMuMax3} and \onlinecite{leliaert2018fast} for a recent review). The local free energy density $\mathcal{E}$ is related to the magnetization $\textbf{M}(x,y)=\Msat\textbf{m}(x,y)$, where $\Msat$ is the saturation magnetization and $\left|\textbf{m}\right|=1$. We consider the free energy resulting from the following magnetic interactions: exchange interaction, perpendicular anisotropy, DMI, Zeeman interaction and demagnetization. We approximate the demagnetization energy by using an effective anisotropy $\Keff=K-\frac{1}{2}\mu_0 \Msat^2$, with $K$ the perpendicular magnetic anisotropy and $\mu_0$ the vacuum permeability, which is justified for the case of ultrathin ferromagnetic films~\cite{coey2010magnetism}. The expressions for the resultant energy-density terms are
\begin{eqnarray}
&&\mathcal{E}_{\text{ex}}=A_{\text{ex}}\left[ \left(\frac{\partial\textbf{m}}{\partial x}\right)^2 +\left(\frac{\partial\textbf{m}}{\partial y}\right)^2\right],\nonumber\\
&&\mathcal{E}_{\text{anis}}=K_{\text{eff}}(1-m_z^2),\nonumber\\
&&\mathcal{E}_{\text{DMI}}=-D\left[m_x\frac{\partial m_z}{\partial x}-m_z\frac{\partial m_x}{\partial x}
+m_y\frac{\partial m_z}{\partial y} -m_z\frac{\partial m_y}{\partial y}\right],\nonumber\\
&&\mathcal{E}_{\text{Zeeman}}=-\Msat\textbf{B}\cdot\textbf{m}.\nonumber
\end{eqnarray}
Our sample is an ultrathin ferromagnetic film with perpendicular magnetic anisotropy, with DMI induced by adjacent heavy metal layer with strong spin-orbit coupling. We consider the following parameters:
saturation magnetization $\Msat=580$~kAm$^{-1}$, exchange stiffness $A_{\text{ex}}=15$~pJm$^{-1}$, and perpendicular anisotropy $K=0.8 $~MJm$^{-3}$ ($\Keff=0.6$~MJm$^{-3}$), stemming from the experimental results on Co/Pt systems \cite{metaxas2007creep,Sampaio2013NucleationNanostructures}. The used values of the DMI constant, $D$, will be specified in the sections below, for what is useful to define the critical DMI strength $D_c=4\sqrt{A_{\text{ex}} K_{\text{eff}}}/\pi$ above which spin-cycloids become the ground-state in the ferromagnetic sample\cite{Rohart2013SkyrmionInteraction}. $\textbf{B}$ represents the external magnetic field, which in our case will be the vortex stray field. For all simulations, we consider a system discretized into cells of size $1\times1\times0.4$ nm$^3$, with $d=0.4$~nm the thickness of the FM film. In mumax$^3$ the dynamics of the magnetization is governed by the Landau-Lifshitz-Gilbert (LLG) equation 
\begin{equation}
    \frac{d\textbf{m}}{dt}=\frac{\gamma}{1+\alpha^2}\left(\textbf{m}\times\textbf{H}_{\text{eff}}+\alpha\left[\textbf{m}\times(\textbf{m}\times\textbf{H}_{\text{eff}})\right]\right), 
\end{equation} 
where $\gamma=1.7595\times10^{11}$~AmN$^{-1}$s$^{-1}$ is the gyromagnetic ratio and $\alpha$ is the Gilbert damping factor.
In this work we consider $\alpha=0.02$ and $0.3$, representing, respectively, the low and high damping regimes of the FM material. $\textbf{H}_{\text{eff}}$ is the effective magnetic field given by the functional derivative of the free energy $E=\int(\mathcal{E}_{\text{ex}}+\mathcal{E}_{\text{anis}}+\mathcal{E}_{\text{DMI}}+\mathcal{E}_{\text{Zeeman}})dV$ with respect to the magnetization: $\textbf{H}_{\text{eff}}=-\frac{1}{\mu_0 \Msat}\delta E/\delta\textbf{m}$. 

\subsection{Equation of motion for the center-of-mass of the skyrmion}\label{secIIc}

Thiele equation describes the dynamics of the center-of-mass of the skyrmion by assuming a rigid body motion of the spin texture \cite{Tomasello2014AMemories,Zhang2016MagneticEffect,Jiang2017DirectEffect,menezes2019deflection}. For the case of in-plane applied current the Thiele equation reads\cite{thiele1973steady}
\begin{equation}
    \textbf{G}\times(\bm{\nu}-\Dot{\textbf{r}}_\text{sk})+\bm{\mathcal{D}}(\beta\bm{\nu}-\alpha\Dot{\textbf{r}}_\text{sk})-\nabla V=0, \label{eq.thiele}
\end{equation}
where $\textbf{G}=\mathcal{G}\hat{z}=4\pi Q(d\Msat/\gamma)\hat{z}$ is the gyromagnetic coupling vector, with $Q$ the skyrmion number (in all simulations we consider $Q=-1$); $\Dot{\textbf{r}}_\text{sk}=\Dot{x}_\text{sk}\hat{x}+\Dot{y}_\text{sk}\hat{y}$ is the skyrmion drift velocity; $V$ is the potential induced by the vortex field; $\bm{\nu}=\nu_x\hat{x}+\nu_y\hat{y}$ is associated to the velocity of the conduction electrons in the spin-polarized current, and $\bm{\mathcal{D}}$ represents the dissipative tensor, with components $\mathcal{D}_{ij}=(d\Msat/\gamma)\int d^2r\partial_i\textbf{m}\cdot\partial_j\textbf{m}=\mathcal{D}\delta_{ij}$ (see Appendix B). Eq.~(\ref{eq.thiele}) can be separated into its two components, which yields
\begin{eqnarray}
&&\Dot{x}_\text{sk}=\frac{1}{\sigma^2_{\alpha\alpha}}\left[\sigma^2_{\alpha\beta}\nu_x +\mathcal{D}\mathcal{G}(\beta-\alpha)\nu_y + \alpha\mathcal{D}F_\text{sv}^{x} + \mathcal{G}F_\text{sv}^{y}\right],\nonumber\\
&&\Dot{y}_\text{sk}=\frac{1}{\alpha\mathcal{D}}\left[\mathcal{G}(\nu_x-\Dot{x}_\text{sk})+\beta\mathcal{D}\nu_y+ F_\text{sv}^{y}\right],\label{eq.xy}
\end{eqnarray}
where $\sigma_{ab}=\sqrt{\mathcal{G}^2+ab\mathcal{D}^2}$, $F_\text{sv}^{x}=-\partial V/\partial x$, and $F_\text{sv}^{y}=-\partial V/\partial y$.

\section{Static properties of the hybrid system}\label{Sec.III}

\subsection{Effects of the vortex field on the uniform ferromagnetic state}\label{S.III.A}

\begin{figure*}[t]
\centering
\includegraphics[width=\linewidth]{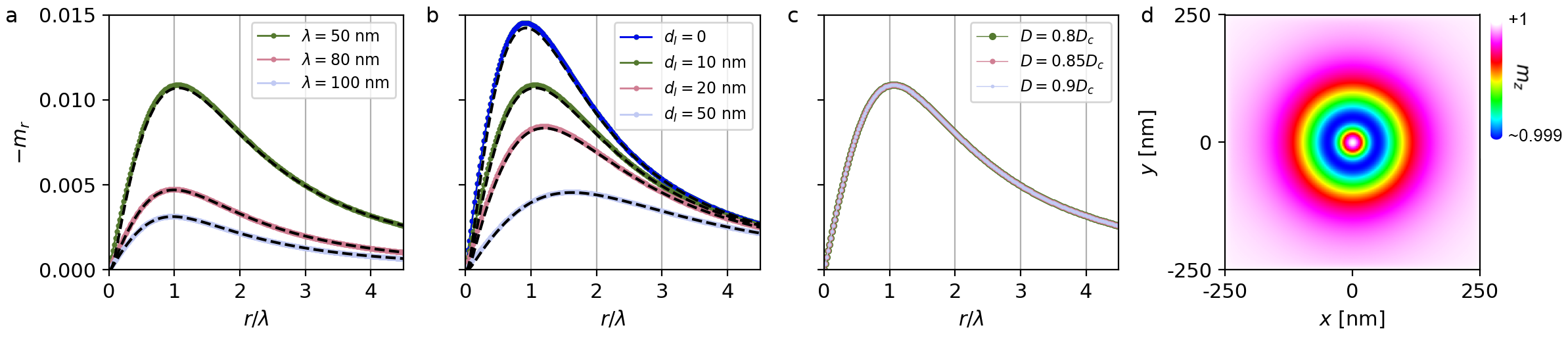}
\caption{Canting induced in the uniform ferromagnetic state of the FM film due to the stray field of the nearby superconducting vortex, as a function of the distance from the vortex core. (a) For different values of penetration depth $\lambda$ of the superconductor, with $d_I=10$~nm and $D=0.8D_c$ fixed. (b) For different values of $d_I$, with $\lambda=50$~nm and $D=0.8D_c$ fixed. (c) For different values of $D$, with $\lambda=50$~nm and $d_I=10$~nm fixed. Dashed lines indicate the corresponding magnetization profiles obtained analytically using Eq. (\ref{Eq.theta}). (d) Contourplot of the $z$-component of the magnetization in the vicinity of the vortex core (centered at $(x,y)=(0,0)$), for $\lambda=50$~nm, $d=10$~nm, and $D=0.8D_c$.}
\label{fig2}
\end{figure*} 

Let us first consider the effects of the magnetic field of the vortex in the superconductor to the uniform ferromagnetic state in the adjacent magnetic film. Fig.~\ref{fig2} shows the magnetization profile obtained from micromagnetic simulations of a ferromagnetic film in the presence of the stray field of a single vortex  in a thick superconducting film, for different values of the penetration depth $\lambda$ of the superconductor, thickness of the insulating layer $d_I$, and DMI strength $D$. The polarity of the magnetic field of the vortex is taken to be negative (pointing along the $-\hat{z}$ direction).

Note that for small values of $\lambda$, where the magnetic flux  induced by the vortex is more localized, the corresponding canting of the magnetization in the FM film is more pronounced. Also notice that, for the parameters considered in this work, the presence of the superconducting vortex does affect the ferromagnetic ground state, but it is not sufficient to nucleate a skyrmion as e.g. considered in Ref. \onlinecite{del2015imprinting}. In fact, assuming weak variations of the local spin tilt angle $\theta$, the magnetization profile induced by the stray field of the vortex can be calculated by considering the micromagnetic energy density in polar coordinates \cite{rohart2013skyrmion}
\begin{equation}
\begin{aligned}
    E^{2D}[\theta(r)]=&2\pi\int_0^{\infty}\left[ A\left( \frac{d\theta}{dr}\right)^2+A\frac{\sin^2\theta}{r^2}\right.\\
    &-D\left(\frac{d\theta}{dr}+\frac{\cos\theta\sin\theta}{r}\right)\\
    &\left.+K_{\text{eff}}\sin^2\theta-M_s B_r\sin\theta-M_s B_z\cos\theta \vphantom{\left( \frac{d\theta}{dr}\right)^2}\right]rdr,\\
    &=\int_0^{\infty}\mathcal{E}\left(\theta,\frac{d\theta}{dr},r\right)dr,
    \label{Eq.E}
\end{aligned}
\end{equation}
where we assumed $\textbf{m}=\sin\theta\hat{r}+\cos\theta\hat{z}$, and $\textbf{B}_v=B_r\hat{r}+B_z\hat{z}$ is the stray field of a vortex located at $r=0$. In the limit of weak variations of the angle $\theta$ ($\frac{d\theta}{dr}\ll1$ and $\theta\ll1$), the energy density can be rewritten as
\begin{equation}
\begin{aligned}
    \mathcal{E}&\left(\theta,\theta',r\right)=2\pi r\left[-D\theta'+\theta^2\left( \frac{A}{r^2}+K_{\text{eff}}+\frac{M_s B_z}{2}\right)\right.\\
    &\left.+\theta\left(-\frac{D}{r}-M_sB_r\right)-M_sB_z+\mathcal{O}(\theta^3)+\mathcal{O}(\theta'^2) \right],
\end{aligned}
\end{equation}
where $\theta'=d\theta/dr$. The Euler-Lagrange equation
\begin{equation}
    \frac{\partial\mathcal{E}}{\partial\theta}-\frac{d}{dr}\left(\frac{\partial\mathcal{E}}{\partial\theta'}\right)=0
\end{equation}
minimizes the energy functional and yields the following expression for the magnetization profile:
\begin{equation}
    \theta(r)=\frac{B_r(r)M_s}{\frac{2A}{r^2}+2K_{\text{eff}}+B_z(r)M_s}.
    \label{Eq.theta}
\end{equation}
Fig. \ref{fig2}(a,b) shows that the above expression (dashed lines) nicely agrees with the magnetization profile obtained in the micromagnetic simulations. Fig. \ref{fig2}(c) shows that, as suggested by Eq. (\ref{Eq.theta}), $\theta(r)$ does not depend on the DMI parameter. Notice that this expression is valid for any radial field such as that created by superconducting vortices, magnetic dots, or nearby magnetic tips, provided that the uniform magnetic order is only weakly perturbed. It will be useful now to define the radius of maximal canting, $r_{\theta}^{\text{max}}$, as given by $\theta(r_{\theta}^{\text{max}})=\text{max}[\theta(r)]$. For the case of $d_I=10~\text{nm}$ and $\lambda=50~\text{nm}$, we find $r_{\theta}^{\text{max}}\approx\lambda$. From here on, $d_I=10~\text{nm}$ will be used in all remaining  calculations, unless stated otherwise. 

\begin{figure}[b]
\centering
\includegraphics[width=7cm]{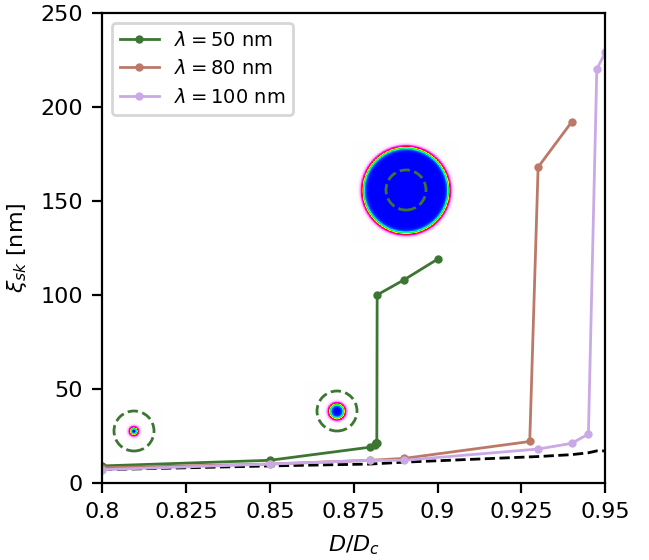}
\caption{Skyrmion radius when on top of a superconducting vortex, extracted from the micromagnetic simulations, as a function of the DMI strength $D$. Dashed line shows the skyrmion size in the absence of an external magnetic field. The insets show the $z$ component of the magnetization for $\lambda=50$~nm, where dashed circles represent $r=r_{\theta}^{\text{max}}$ i.e. area where vortex core has strongest influence on the ferromagnetic state.}
\label{figD}
\end{figure}
\begin{figure*}[t]
\centering
\includegraphics[width=\linewidth]{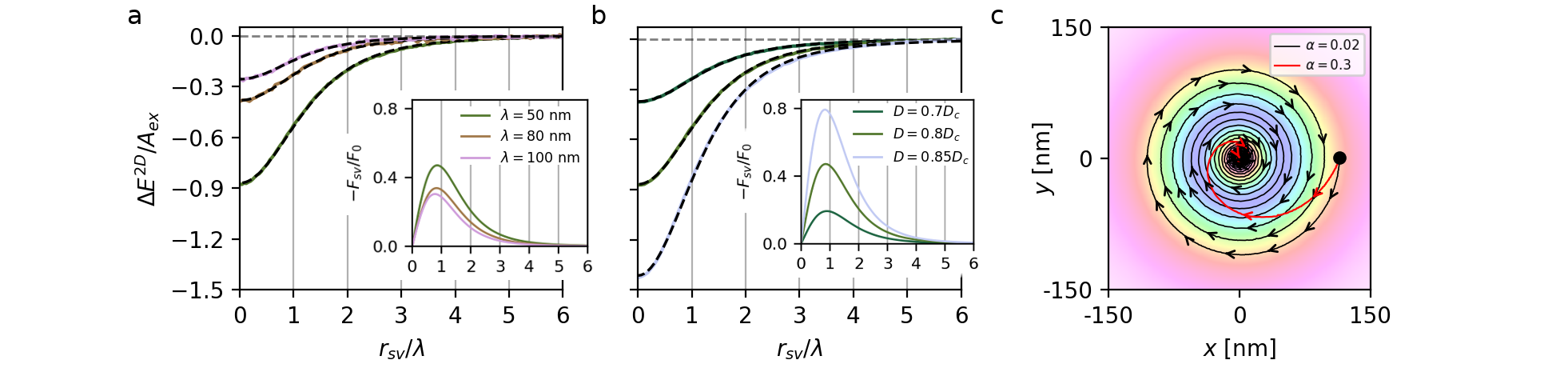}
\caption{Skyrmion-vortex interaction energy calculated in the micromagnetic simulations as a function of the distance between the skyrmion and vortex cores for (a) different values of $\lambda$ and fixed $D=0.8D_c$, and (b) different values of $D$ and fixed $\lambda=50$~nm. The curves fitted by Eq. (\ref{Eq.fit}) are shown as dashed lines. Insets show the corresponding interaction force calculated as the derivative of the fitted curves, where $F_0=dA_\text{ex}/\lambda_0=0.12$~pN, with $\lambda_0=50$~nm. (c) Trajectory of a skyrmion dynamics in the presence of the vortex field, for $\alpha=0.02$ and $0.3$, $\lambda=50$~nm, and $D=0.8D_c$. Black dot indicates the initial position of the skyrmion and the arrows the center-of-mass trajectories. Background colors show the $z$ component of the magnetization induced by the vortex in absence of a skyrmion, as shown in Fig. \ref{fig2}(d).}
\label{figE}
\end{figure*}

\subsection{Effects of the vortex field on the skyrmion size}\label{S.III.B}

The stray field of the vortex can affect the skyrmion size by favoring the rotation of the spin texture in the direction of the flux lines, where the competition with other magnetic interactions is relevant. For simplicity, we will only consider variation of the DMI strength and fix all the remaining parameters of the ferromagnetic material. By increasing the DMI strength one favors the rotation of the magnetization at short length scales and reduces the energy barrier for the vortex field to flip the spins along its direction, resulting in an increase of the skyrmion size. Fig.~\ref{figD} shows how the skyrmion size, calculated by micromagnetic simulations, is affected by the vortex field, for different values of $D$ and $\lambda$, where skyrmion and vortex are on top of each other and concentric. For each $\lambda$, if $D\leq D^{\ast}_{\lambda}$, the skyrmion radius $\xi_{\text{sk}}$ is confined in a region $\xi_{\text{sk}}<r_{\theta}^{\text{max}}$, and increases its size abruptly to $\xi_{\text{sk}}>r_{\theta}^{\text{max}}$ when $D$ exceeds $D^{\ast}_{\lambda}$. The threshold state, where $\xi_{\text{sk}}\approx r_{\theta}^{\text{max}}$, is unstable. From the simulations we calculate $D^{\ast}_{\lambda}\approx 0.882D_c$, $0.9275D_c$ and $0.945D_c$ for $\lambda=50$, $80$ and $100$~nm respectively. Notice that there is a range of DMI ($D<0.85D_c$ for all considered $\lambda$'s) where the skyrmion size is weakly affected by the presence of the superconducting vortex and $\xi_{\text{sk}}$ approximately corresponds to the skyrmion size in the absence of any magnetic field (dashed line in Fig.~\ref{figD}). In this case, the interaction energy is dominated by the difference in Zeeman energy due to the presence of the vortex stray field. Nevertheless, the other terms are highly sensitive to the change in the skymion shape and thereby give a non-negligible contribution to the total vortex-skyrmion interaction energy (see Appendix~\ref{AppB}).

\subsection{Skyrmion-vortex interaction}\label{S.III.C}

As shown in the previous section, the skyrmion-vortex interaction is stronger when the domain wall of the skyrmion faces the maximal background canting, i.e., when the skyrmion core is at a distance $r_{c}\approx |r_{\theta}^{\text{max}}-\xi_{\text{sk}}|$ from the vortex core. To numerically calculate the spatial profile of the interaction energy between the skyrmion and the superconducting vortex, we relax the magnetization in the micromagnetic simulation for different positions of the vortex stray field, while keeping
the magnetic moment at the center of the skyrmion fixed, at a fixed location. This approach is similar to the method used in Refs. \onlinecite{muller2015capturing,lin2013particle,mulkers2017effects} to calculate the interaction of the skyrmion with holes, sample edges, or material defects. We consider the case of $D\leq0.85D_c$, where the skyrmion profile is weakly perturbed by the presence of the vortex field (see Fig. \ref{figD}). In such a situation we are sure that the fixed core will indeed represent the center of mass of the skyrmion after relaxing the magnetization. Fig. \ref{figE}(a,b) shows the interaction energy calculated in these simulations, as a function of the distance between the skyrmion and vortex cores, $r_{\text{sv}}$, for different values of $\lambda$ and DMI strength $D$. Notice that the obtained energy profile can be fitted numerically by the expression 
\begin{equation}
    E=\frac{a}{(r_{\text{sv}}^2+b\lambda^2)^2},
    \label{Eq.fit}
\end{equation}
with $a$ and $b$ the fitting parameters. The fitted curves are shown as dashed lines in Fig. \ref{figE}(a,b). Insets show the corresponding interaction forces derived from Eq.~(\ref{Eq.fit}). 

For further analysis, we fix $\lambda=50$~nm and $D=0.8D_c$ in the simulations, unless specified otherwise. 

\section{Skyrmion dynamics in the presence of a superconducting vortex}\label{Sec.IIIb}

\begin{figure*}[t]
\centering
\includegraphics[width=0.9\linewidth]{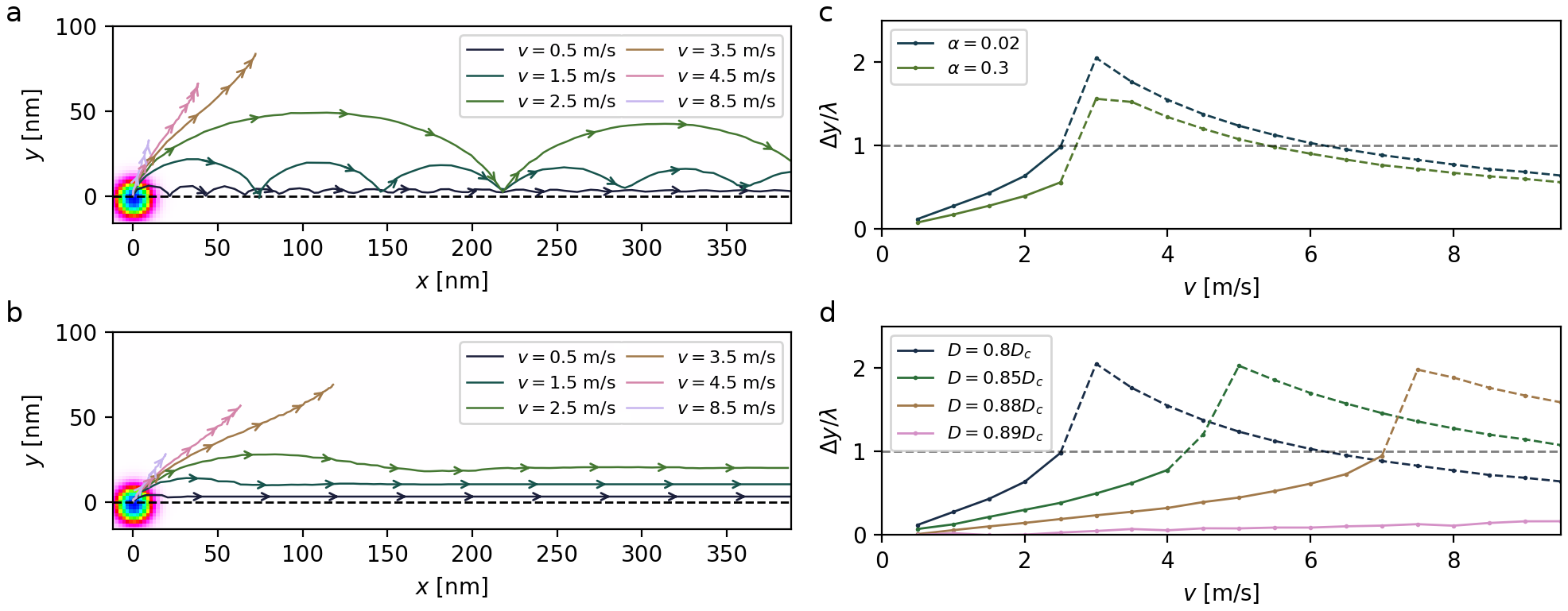}
\caption{Trajectories of the center of mass of the skyrmion calculated in the micromagnetic simulations for different values of the driven vortex velocity $v$, with damping factor $\alpha=0.02$ (a), or $\alpha=0.3$ (b). (c) Maximal amplitude of the cycloidal trajectory as a function of $v$, for different values of $\alpha$, with fixed $D=0.8D_c$. (d) Maximal amplitude of the cycloidal trajectory as a function of $v$ for different values of the DMI strength and $\alpha=0.02$ fixed. Transition from solid to dashed line indicates the escape velocity.}
\label{fig3}
\end{figure*}

\subsection{Vortex at rest}\label{sec.dynamics-rest}\label{S.IV.A}

We start by describing the motion of the skyrmion induced by the interaction with a pinned vortex in the superconductor, without any other applied drive. Fig.~\ref{figE}(c) shows the center-of-mass trajectories of the skyrmion in the presence of the vortex field, calculated in the micromagnetic simulations with damping parameter $\alpha=0.02$ or $0.3$, where the vortex position is fixed at the center of the simulation box and the skyrmion is initialized at a distance $r_{\text{sv}}=2.4\lambda$ from the vortex core. As shown in energetic considerations of the previous section, the skyrmion is indeed attracted to the vortex core. A deflection in the azimuthal $\varphi$ direction is induced by the Magnus force ($\varphi$ is the angular cylindrical coordinate with origin at the vortex core position), and the skyrmion follows a spiral trajectory towards the center of the vortex. Damping factor $\alpha$ controls the magnitude of the Magnus force and consequently the shape of the spiral trajectory. Similar trajectories are observed, e.g., for the skyrmion approaching a pinning center\cite{liu2013mechanism,lin2013particle}.


\subsection{Vortex at constant speed}\label{sec.dynamics-const-speed}\label{S.IV.B}

Let us next consider that a uniform current density, $j_{SC}$, is applied into a conventional superconducting material. The current induces a Lorentz force $\textbf{F}_\text{L}= d_\text{SC}\phi_0\textbf{j}_\text{SC}\times\hat{z}$, which acts on the vortex core, thus forcing the vortex to move and, consequently, inducing the skyrmion motion as well. As a first approximation, in this section we neglect the effects of the skyrmion to the  vortex  motion  and  consider the vortex to move straight along the Lorentz force at a constant speed given by $v=F_L/\eta$, where $\eta$ is the vortex viscous drag coefficient.
As we shall discuss in more detail in Sec.˜~\ref{sec.dynamics-CurrSC}, this is a good approximation only when both the driving force and the viscous drag acting upon the vortex are much stronger than the vortex-skyrmion force. 

We performed micromagnetic simulations initializing the magnetic skyrmion concentric to the vortex core and then moving the vortex field, in a rigid body motion, along the $+\hat{x}$ direction, with constant velocity $v$.  Fig.~\ref{fig3}(a) shows the corresponding trajectories of the skyrmion for different values of $v$ and for damping constant $\alpha=0.02$. The skyrmion moves in cycloidal arcs created by the competition between the movement along the $\hat{x}$ direction imposed by the driven vortex and the deflection along the $\varphi$ direction with respect to the vortex. The maximal amplitude of the cycloidal trajectory is approximately $\lambda$, which coincides with the maximal canting region defined by $r_\theta^\text{max}$. For $v$ higher than an escape velocity, $v_c$, the skyrmion crosses the $r=r_\theta^\text{max}$ region and escapes from the confinement by the vortex field. The maximal amplitude $\Delta y$ of the skyrmion arc trajectory as a function of the vortex velocity is shown in Fig.~\ref{fig3}(c) for $\alpha=0.02$ and $0.3$, with $D=0.8D_c$ fixed, and in Fig.~\ref{fig3}(d) for different values of $D$ and $\alpha=0.02$ fixed. In the latter case, for $D>D_{\lambda}^{\ast}$ one has $\xi_\text{sk}>r_{\theta}^{\text{max}}$ and the skyrmion trajectory no longer presents periodic arcs during the motion. Notice that the escape velocity does not change considerably by changing from low to high damping regime, however it strongly depends on the DMI parameter, as expected from the interaction force in Sec.~\ref{S.III.C}. 

Similar cycloidal motion has been observed in Ref. \onlinecite{wang2017manipulating} for a moving magnetic field, where the authors stated that the skyrmion follows a periodic motion. However, notice from Fig.~\ref{fig3}(a) that the amplitude of the cycloidal arcs decreases as the skyrmion moves further. In fact, by increasing the damping factor the dynamics changes from underdamped to overdamped motion, as show in Fig.~\ref{fig3}(b) for $\alpha=0.3$. Therefore, the cycloidal motion is a transient motion, after which the trajectories converge to a situation where the skyrmion moves along with the vortex, keeping a constant nonzero distance from the vortex core position (thick dashed lines in Figs.~\ref{fig3}(a,b)). This indicates that the vortex core is no longer the minimal energy position for the skyrmion in the dynamical system as it is for the system with a stationary vortex ($v=0$). 
\begin{figure*}[t]
\centering
\includegraphics[width=\linewidth]{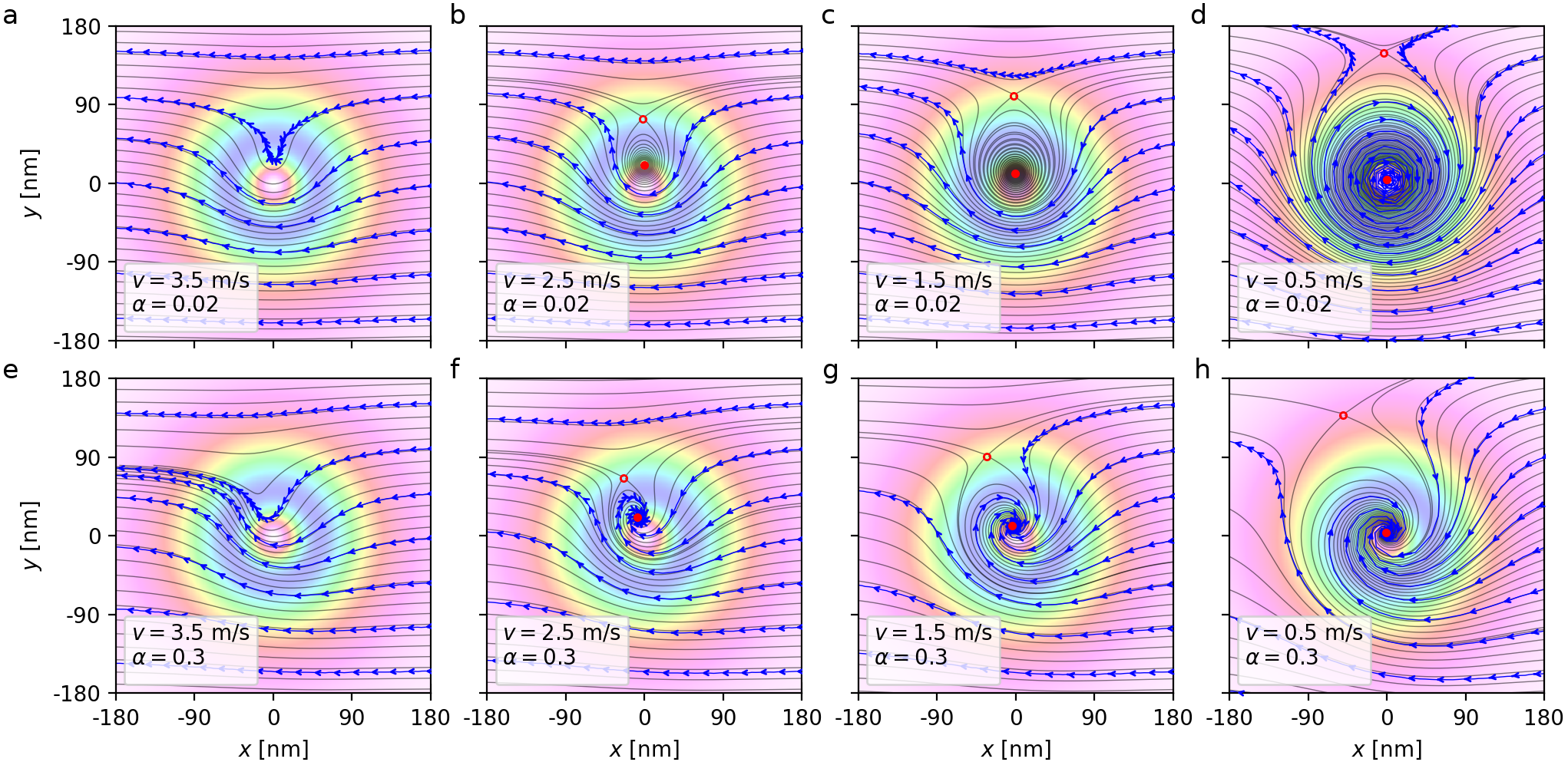}
\caption{Arrows show the skyrmion trajectories calculated in the micromagnetic simulations for different values of the vortex velocity, $v$, with $\alpha=0.02$ (a-d) and $0.3$ (e-h), plotted in the frame of reference of the moving vortex. Thin lines are the corresponding trajectories calculated from the Thiele equation. Dots show the fixed points, where open dots indicate saddle points and closed dots represent stable spiral points. Background colors show the $z$ component of the magnetization induced in the absence of a skyrmion, as shown in Fig. \ref{fig2}(d).}
\label{fig7}
\end{figure*}

The above behavior is better understood in the frame of reference of the moving superconducting vortex. Fig. \ref{fig7} shows the trajectories (indicated by arrows) of the center-of-mass of the skyrmion calculated in the micromagnetic simulations for different values of the vortex velocity, $v$, for $\alpha=0.02$ or $0.3$, in the frame of reference of the moving vortex. Each trajectory corresponds to a different initial position of the skyrmion with respect to the vortex core position. Notice that each point of the coordinate space 
belongs to a unique and well defined trajectory which converges to a fixed point or to infinity. Such dynamical behavior can be described in the Thiele formalism by the equation of motion for the center of mass of the skyrmion (see Sec. \ref{secIIc}). In this frame of reference, the magnetic system is moving with velocity $-v\hat{x}$ with respect to the vortex and the skyrmion dynamics can be equivalently described by the situation where a spin-polarized current is applied into the ferromagnetic film along the $\hat{x}$ direction in the particular case where $\alpha=\beta$, and the vortex is at rest. In this case, in regions far from the vortex core, where $\partial V/\partial r=0$, the skyrmion velocity is given by $\dot{\bm{r}}_\text{sk}=\bm{\nu}=-v\hat{x}$. As the skyrmion approaches the vortex, its trajectory can be attracted by one of the fixed points, \{$\textbf{r}^{\ast}$\}, which can be calculated by setting $\Dot{r}^{\ast}_\text{sk}=0$ in Eq. (\ref{eq.xy}). In cylindrical coordinates,  
\begin{subequations}
\begin{align}
   &\varphi^{\ast} =\arctan\left(\frac{\mathcal{G}}{\alpha\mathcal{D}}\right)+n\pi\quad(\text{for $v\neq0$}),\label{Eq.8a}\\
   &\frac{\partial V}{\partial r}\bigg|_{r^{\ast}} =\pm\sigma_{\alpha\alpha} v, \label{Eq.8b}
\end{align}
\end{subequations}
where $n=0,1,2,...$ represent the solutions for both vortex ($-\frac{\partial V}{\partial r}<0$) and antivortex ($-\frac{\partial V}{\partial r}>0$) if $n$ is odd or even respectively. Comparing Eq.~(\ref{Eq.8b}) 
with the skyrmion-vortex interaction force in Fig.~\ref{figE} (insets), there can be 0, 1 or 2 fixed points for $v>v_c$, $v=v_c$, and $v<v_c$ respectively, where $v_c=F_\text{sv}^{\text{max}}/\sigma_{\alpha\alpha}$ is the critical velocity. The stability of the fixed points can be calculated either analytically, by the linearization of the equation of motion near the fixed points, or numerically, by iterating Eqs. (\ref{eq.xy}) in discrete steps of time. Here we apply the second approach, where we take $\alpha=\beta$, $\nu_x=-v$, $\nu_y=0$, and force $\textbf{F}_\text{sv}$ as calculated in Sec.~\ref{S.III.C}. The value of $\mathcal{D}$ was calculated as explained in Appendix B. The corresponding trajectories and fixed points calculated from the Thiele equation are shown in Fig.~\ref{fig7} as lines and dots, respectively, with the open dots representing saddle points and the closed dots representing stable spirals. Notice that Thiele approach is in good agreement with the micromagnetic simulations for the considered parameters. Also notice that with increasing the vortex velocity the fixed points approach until they annihilate around the region of maximal background canting due to the vortex field.       
\begin{figure*}[t]
\centering
\includegraphics[width=\linewidth]{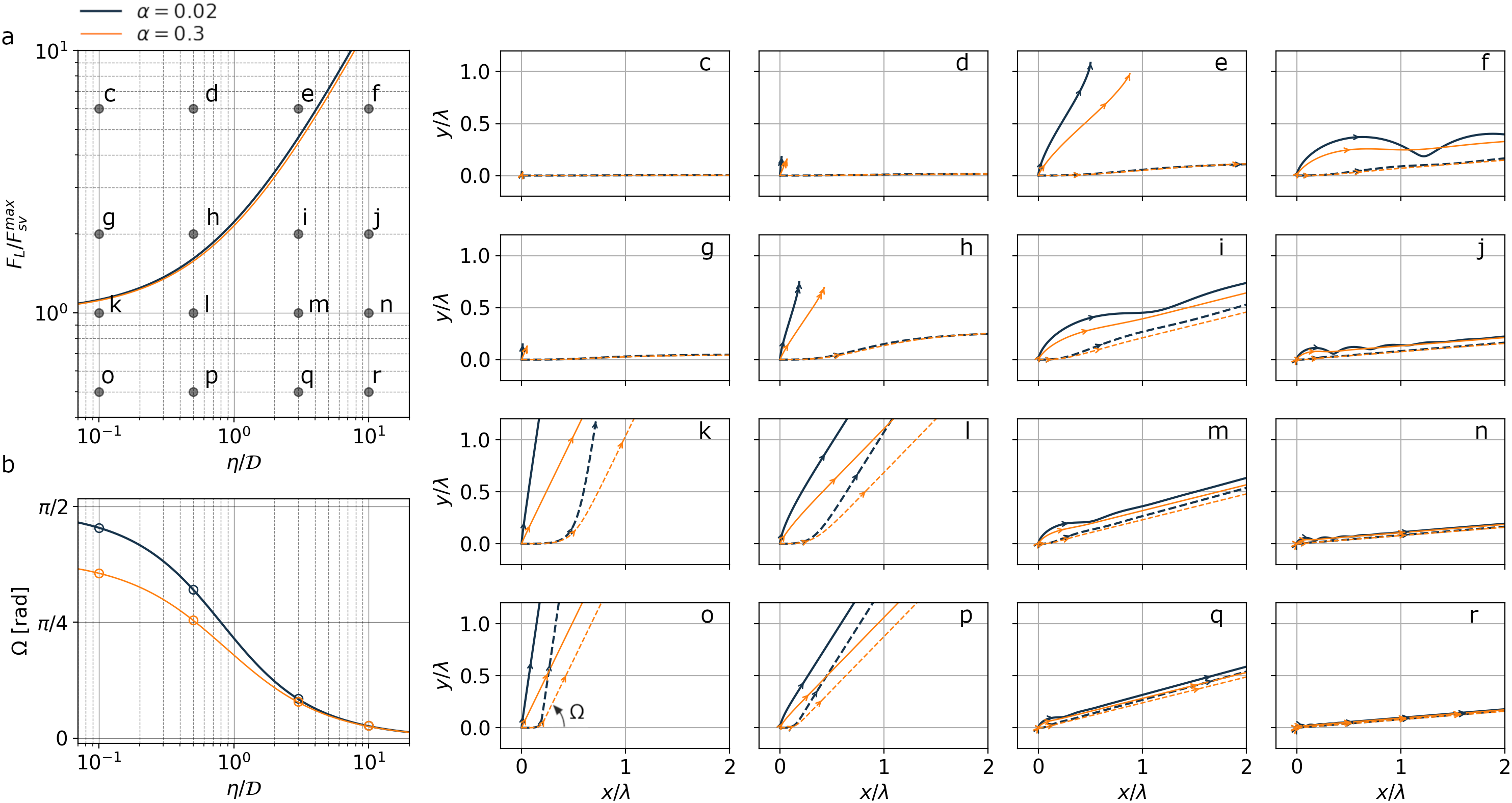}
\caption{(a) The critical force calculated from Eq. (\ref{Eq.flc}), for $\alpha=0.02$ and $\alpha=0.3$. The labeled points represent the parameters ($F_\text{L}$,$\eta$) considered in the simulations. (b) Resultant direction (angle $\Omega$) of the SVP motion with respect to the $\hat{x}$ direction. The open dots indicate the angle calculated from the simulations and solid lines are given by Eq. (\ref{eq.omega}). (c-r) Molecular dynamics simulations for labeled choices of parameters in (a), with the vortex trajectories represented by dashed lines and the skyrmion trajectories by solid lines.}
\label{fig10}
\end{figure*}

\subsection{Applying current into the superconductor} 
\label{sec.dynamics-CurrSC}

As a next step in the analysis, we introduce the feedback effect of the skyrmion dynamics on the driven vortex dynamics by taking into account the vortex-skyrmion interaction in the vortex equation of motion. 
For simplicity, here we consider the limit $d_\text{SC}\ll\lambda$, where the currents in the superconducting film can be averaged over the film thickness and the vortex-core dynamics can be approximated as one of a point particle.

The Bardeen-Stephen equation~ \cite{bardeen1965theory} describes the overdamped motion of the vortex core, with terminal velocity $\Dot{\textbf{r}}_{\text{v}}$ given by the force balance: $\eta\Dot{\textbf{r}}_{\text{v}}=\textbf{F}$, where $\eta$ is a viscosity coefficient and $\textbf{F}$ comprises all other forces acting on the vortex core. In this work we neglect the effects of vortex pinning in the superconductor, as well as the intrinsic vortex Hall effect (negligible outside the superclean limit\cite{parks1995phase,gor1975vortex}), and write the force acting on the vortex core as $\textbf{F}=\textbf{F}_\text{L}-\textbf{F}_\text{sv}$, with $\textbf{F}_\text{L}= d_\text{SC}\phi_0\textbf{j}_\text{SC}\times\hat{z}$ the Lorentz force due to the current density $\textbf{j}_\text{SC}$ applied into the superconductor and $\textbf{F}_\text{sv}$ the skyrmion-vortex interaction force. Therefore, for the case of $\textbf{F}_\text{L}=F_\text{L}\hat{x}$, the equation of motion for the vortex core can be separated as 
\begin{eqnarray}
   &&\Dot{x}_\text{v}=\frac{1}{\eta}(F_\text{L}-F_\text{sv}^x),\nonumber\\
   &&\Dot{y}_\text{v}=-\frac{1}{\eta}F_\text{sv}^y.\label{Eq.BS} 
\end{eqnarray}

The threshold current applied into the superconductor that breaks the skyrmion-vortex pair (SVP) is reached when the vortex attains the critical velocity,   i.e, $\eta v_c=|\textbf{F}|=\sqrt{(F_\text{L}-F_\text{sv}^{x})^2+(F_\text{sv}^{y})^2}$. The critical value of $F_\text{L}$ then reads
\begin{equation}
\begin{aligned}
    F_\text{L}^c=\text{max}\left[|F_\text{sv}^{x}|+\sqrt{\left(\eta v_c\right)^2-(F_\text{sv}^{y})^2}\right]\label{eqFC}.
\end{aligned}
\end{equation}
Here $v_c=F_\text{sv}^\text{max}/\sigma_{\alpha\alpha}$, and we obtain
\begin{equation}
    F_\text{L}^c=\left(1+\frac{\eta}{\sigma_{\alpha\alpha}}\right)F_\text{sv}^\text{max}\label{Eq.flc}.
\end{equation}
Above this value, the fixed points of our dynamical system annihilate, and the skyrmion is left behind when the vortex moves. On the other hand, for $F_\text{L}<F_\text{L}^c$, the SVP remains bound, and after a transient oscillatory motion, the pair reaches a steady state (the dynamical system finds the stable fixed point), where the skyrmion and vortex move with the same velocity, i.e., $\Dot{x}_\text{sk}=\Dot{x}_\text{v}=v_x$ and $\Dot{y}_\text{sk}=\Dot{y}_\text{v}=v_y$, with $v_x$ and $v_y$ constant. By substituting that into Eqs. (\ref{eq.xy}) and (\ref{Eq.BS}), one can calculate the resulting net angle (direction) of the SVP motion with respect to the $\hat{x}$ direction as $\Omega\equiv\arctan(v_y/v_x)$. For the case where there are no currents applied into the ferromagnetic film, i.e, $\nu_x=\nu_y=0$, one obtains
\begin{equation}
    \Omega=\arctan\left(-\frac{\mathcal{G}}{\alpha\mathcal{D}+\eta}\right)\label{eq.omega}.
\end{equation}

In the previous section we have shown that the dynamics of the center-of-mass of the skyrmion, described by the Thiele formalism, is in good agreement with the micromagnetic simulations for the considered range of parameters where the skyrmion size is weakly affected by the vortex field. Therefore, in this section we perform a series of molecular dynamics simulations of the combined skyrmion-vortex system by numerically integrating the coupled Thiele \eqref{eq.xy} and Bardeen-Stephen \eqref{Eq.BS} equations. However, since we are now considering a thin superconducting film, i.e. $d_\text{SC}\ll\lambda$, the monopole approximation is no longer accurate\cite{carneiro2000vortex,thiel2016quantitative} and we numerically integrate Eqs. (\ref{eq0a}) and (\ref{eq0b}) to obtain the vortex stray field. The interaction force is calculated as in Sec.~\ref{S.III.C} (see Appendix \ref{AppB}). For the simulations we consider $\lambda=50$~nm and $d_\text{SC}=10$~nm, however, the results presented in this section can be easily generalized to other values of the parameters of the superconducting film. We initialize the system with the skyrmion and vortex concentric and apply a constant Lorentz force $\textbf{F}_\text{L}=F_\text{L}\hat{x}$ to the vortex, i.e, an uniform current density $\textbf{j}_\text{SC}=-j_\text{SC}\hat{y}$ is applied into the superconductor. Panels (c-r) in Fig.~\ref{fig10} show the trajectories obtained in the simulations, where Eq.~(\ref{Eq.flc}) is used as reference for the considered parameters, as indicated in Fig.~\ref{fig10}(a). Fig.~\ref{fig10}(b) shows that the observed angle of the resultant motion of the SVP agrees with Eq.~(\ref{eq.omega}). Notice that now the skyrmion can experience many different transient motions and follow different directions, depending on the material parameters and Lorentz force. For high values of $\eta$, the dynamical system converges to the one considered in the last section, where $\Omega$ goes to zero and the vortex moves straight along the Lorentz force direction. For the limit of low viscosity of the superconductor and ferromagnet, the SVP motion approaches the direction of the current applied into the superconductor, i.e, perpendicular to the Lorentz force!

 Typical experimental values of the viscous drag coefficient for thin films of conventional superconducting materials are  $\eta/d_\text{sc}\sim10^{-8}\numrange{}{}10^{-6}$~Ns/m$^2$. \cite{parks1995phase,embon2017imaging,alexandre1996coherence} Comparing these values with the skyrmion dissipative-tensor $\mathcal{D}\approx2\times10^{-16}$~Ns/m calculated in Appendix B for the considered FM film, one finds  $\eta/\mathcal{D}\sim \numrange{0.5}{500}$ for a superconducting film of thickness $d_\text{sc}\sim\numrange{10}{100}$~nm. Notice that, once the material has been chosen, the relation $\eta/\mathcal{D}$ can still be tuned by changing the thickness of both FM and SC films, as well as by changing the heavy metal capping layer, which in turn affects the DMI and the skyrmion size. This allows for a high degree of controllability over the angle $\pi/2-\Omega$ between the SVP motion and the current applied into the superconductor, and thereby, over the different dynamical regimes shown in Fig.~\ref{fig10}.

\subsection{Guiding magnetic skyrmions by vortex-screened Hall effect}\label{S.IV.C}

In this section we analyze the full potential for guiding magnetic skyrmions by tuning the skyrmion-vortex Hall effect in FM-SC heterostructures. For that purpose, we now consider that independent currents are applied into both FM and SC films. As in the previous section, if one assumes that after a transient oscillatory motion the SVP reaches the steady dynamic state, where skyrmion and vortex move with the same constant velocity, the angle of the SVP motion with respect to the $\hat{x}$ direction, now with $\nu_x,\nu_y\neq0$, becomes
\begin{equation}
\tan\Omega=\frac{\mathcal{G}}{\alpha\mathcal{D}+\eta}\left[\frac{\Xi_1(\nu_x+\frac{\beta\mathcal{D}}{\eta}\nu_y)}{\Xi_2\nu_x+\Xi_3\nu_y+(\alpha\mathcal{D}+\eta)F_\text{L}}  -1 \right],\\ \label{eq.omega2}
\end{equation}
where
\begin{eqnarray}
&&\Xi_1= \sigma_{\alpha\alpha}^2+2\alpha\mathcal{D}\eta+\eta^2,\nonumber\\
&&\Xi_2= \sigma_{\alpha\beta}^2+\beta\mathcal{D}\eta,\nonumber\\
&&\Xi_3= \mathcal{G}\mathcal{D}(\beta-\alpha-\eta/\mathcal{D}).\nonumber
\end{eqnarray}
The above equation describes the terminal motion of the SVP in a general situation where currents are applied into both FM and SC films. Notice that the direction of the terminal motion does not depend on the strength of the skyrmion-vortex interaction, it depends only on the material parameters and the applied currents. The skyrmion-vortex interaction will nevertheless define the critical forces under which the pair remains connected. Similar expression has been obtained in Ref. \onlinecite{hals2016composite} by a different approach, where Lorentz force due to currents applied into the superconductor was not considered. At this point, we call for attention to three different scenarios in Eq.~(\ref{eq.omega2}). (i) \textit{The current is applied only into the SC film}. In this case we recover Eq.~(\ref{eq.omega}) by substituting $\nu_x=\nu_y=0$ into Eq.~(\ref{eq.omega2}), and $0<\Omega<\pi/2$, as verified in Fig.~\ref{fig10}(b). (ii) \textit{The current is applied only into the FM film}. This case is obtained by choosing $F_\text{L}=0$ in Eq.~(\ref{eq.omega2}), where the case of $\nu_x>0$ and $\nu_y=0$ results in $-\pi/2<\Omega<\Omega_0$, with $\Omega_0=\tan^{-1}[\mathcal{G}\mathcal{D}(\alpha-\beta)/\sigma_{\alpha\beta}^2]$ the skyrmion Hall angle in the absence of the vortex. In other words, the SVP Hall-angle with respect to currents applied into the ferromagnetic film, $\theta_\text{H}^{j_\text{FM}}=\Omega$,  is always greater than  that observed in the absence of superconducting vortices. (iii) \textit{The current is applied into both FM and SC films}. In this case we explore two different situations of the spin-polarized current, $\bm{\nu}\parallel\textbf{F}_\text{L}$ and $\bm{\nu}\perp\textbf{F}_\text{L}$. The Lorentz force, $F_\text{L}^{\ast}$, that compensates the SHE, i.e, that makes the skyrmion move straight along the current direction, is obtained by setting ($\Omega=0$, $\nu_x=\nu$, $\nu_y=0$) and ($\Omega=\pi/2$, $\nu_x=0$, $\nu_y=\nu$) in Eq.~(\ref{eq.omega2}) for $\bm{\nu}\parallel\textbf{F}_\text{L}$ and $\bm{\nu}\perp\textbf{F}_\text{L}$ respectively:
\begin{subequations}
\begin{align}
    F_\text{L}^{\ast}&=\frac{\Xi_1-\Xi_2}{\alpha\mathcal{D}+\eta}\nu,\quad\text{ for }(\bm{\nu}\parallel\textbf{F}_\text{L}), \label{eq17a} \\
    F_\text{L}^{\ast}&=-\frac{\Xi_3}{\alpha\mathcal{D}+\eta}\nu,\quad\text{ for }(\bm{\nu}\perp\textbf{F}_\text{L}).\label{eq17b}
\end{align}
\end{subequations}

\begin{figure}[t]
\centering
\includegraphics[width=8cm]{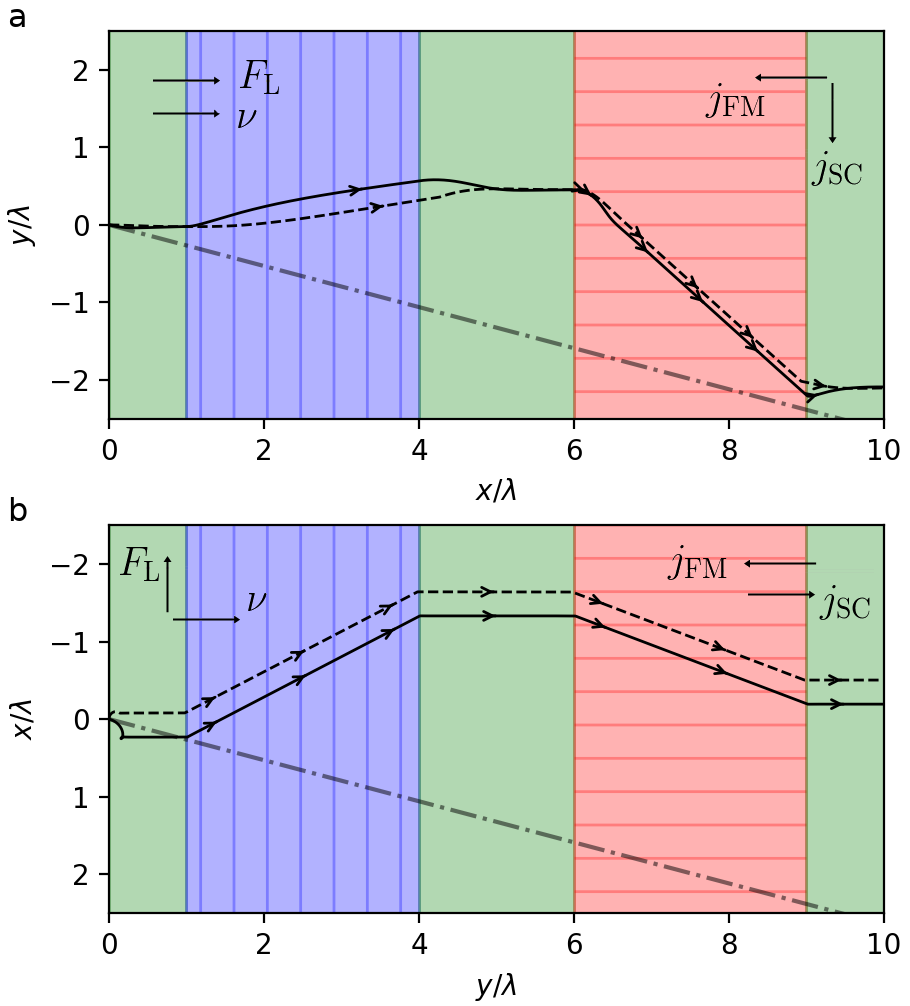}
\caption{Trajectories calculated in the molecular dynamics simulations for (a) $\bm{\nu}\parallel\textbf{F}_\text{L}$, and (b) $\bm{\nu}\perp\textbf{F}_\text{L}$, where dashed and solid lines represent the vortex and skyrmion trajectories respectively, for $F_\text{L}=F_\text{L}^{\ast}$ ((green) solid shaded region), $F_\text{L}=F_\text{L}^{\ast}+\delta F_\text{L}$ ((blue) vertically striped region) and $F_\text{L}=F_\text{L}^{\ast}-\delta F_\text{L}$ ((red) horizontally striped region). The dash-dotted line represents the skyrmion Hall angle in the absence of the vortex. Taken parameters are $\alpha=0.3$, $\beta=\alpha/4$, $\eta=2\mathcal{D}$ and $|\bm{\nu}|=200\nu_0\approx1$~ms$^{-1}$, with $\nu_0\equiv F_{sv}^\text{max}/(\alpha\mathcal{D}+\eta)$. We use $\delta F_\text{L}=160F_\text{sv}^\text{max}$ in (a) and $\delta F_\text{L}=3.2F_\text{sv}^\text{max}$ in (b).}
\label{fig11}
\end{figure}
\begin{figure}[t]
\centering
\includegraphics[width=\linewidth]{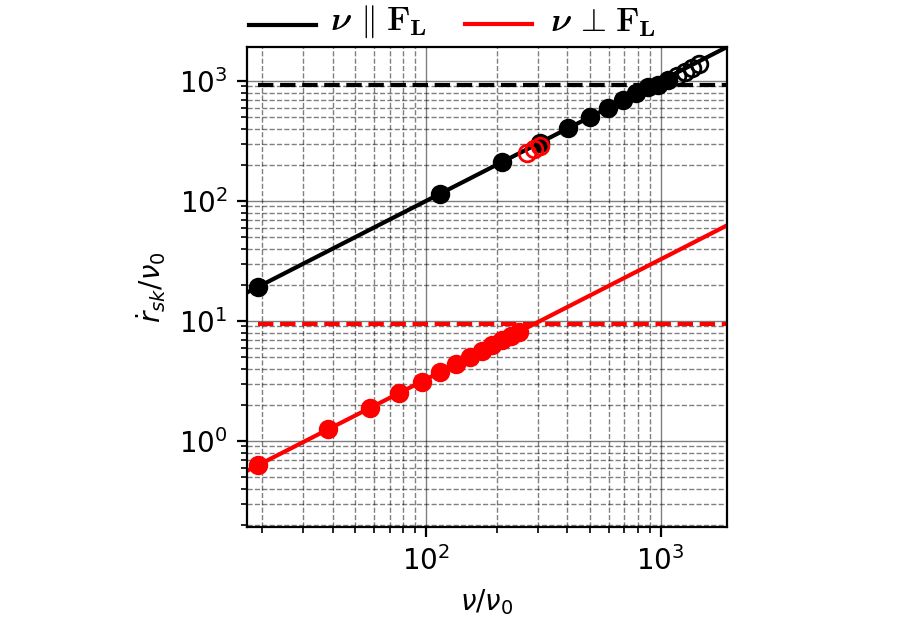}
\caption{Skyrmion terminal velocity as a function of the applied polarized current, for $\bm{\nu}\parallel\textbf{F}_\text{L}$ (black) and $\bm{\nu}\perp\textbf{F}_\text{L}$ (red), with $F_\text{L}$ given by Eqs.~(\ref{eq17a}) and (\ref{eq17b})  so as to compensate the skyrmion Hall effect. Solid lines indicate the expected SVP velocity from Eqs.~(\ref{eqvpa}) and (\ref{eqvpb}). Dots show the results obtained from the simulation, where open dots indicate that the SVP has been broken and the skyrmion motion is no longer aligned with current direction. Dashed lines denote the critical velocities calculated from Eqs.~(\ref{eqvca}) and (\ref{eqvcb}).      }
\label{fig12}
\end{figure}
Figs.~\ref{fig11} (a) and (b) show the trajectories calculated in the molecular dynamics simulations for $\bm{\nu}\parallel\textbf{F}_\text{L}$ and $\bm{\nu}\perp\textbf{F}_\text{L}$ respectively, where we assume the typical values for Co/Pt samples $\alpha=0.3$, $\beta=\alpha/4$, and $\eta=2\mathcal{D}$ for the superconducting film, with $|\bm{\nu}|=200\nu_0\approx1$~ms$^{-1}$, with characteristic velocity $\nu_0\equiv F_{sv}^\text{max}/(\alpha\mathcal{D}+\eta)$. Notice that for $F_\text{L}=F_\text{L}^{\ast}$ (solid shaded (green) regions in Fig.~\ref{fig11}) the SHE is indeed canceled and the SVP moves straight along the current direction. Also notice that by tuning the Lorentz force one can control the direction of motion. By assuming the special cases of Eqs.~(\ref{eq17a}) and (\ref{eq17b}) in the expression for the SVP terminal velocity, one finds
\begin{subequations}
\begin{align}
    v_\text{pair}^{\ast}&=\nu,\quad\text{ for }(\bm{\nu}\parallel\textbf{F}_\text{L})\label{eqvpa},\\
    v_\text{pair}^{\ast}&=\frac{\beta}{\alpha+\eta/\mathcal{D}}\nu,\quad\text{ for }(\bm{\nu}\perp\textbf{F}_\text{L})\label{eqvpb},
\end{align}
\end{subequations}
where $v_\text{pair}^{\ast}$ is the SVP velocity along the direction of applied current. The maximal velocity for which the SVP remains bound together is obtained by substituting Eqs.~(\ref{eq17a}) and (\ref{eq17b}) into Eq.~(\ref{eqFC}), with $\nu$ given by the critical limit of Eqs.~(\ref{eqvpa}) and (\ref{eqvpb}), yielding  
\begin{subequations}
\begin{align}
    v_c^{\ast}&=\frac{F_\text{sv}^\text{max}}{\mathcal{D}(\alpha-\beta)}   ,\quad\text{ for }(\bm{\nu}\parallel\textbf{F}_\text{L})\label{eqvca},\\
    v_c^{\ast}&=\frac{\beta\mathcal{D} F_\text{sv}^\text{max}}{\Xi_3-\beta\mathcal{D}\eta}   ,\quad\text{ for }(\bm{\nu}\perp\textbf{F}_\text{L})\label{eqvcb}.
\end{align}
\end{subequations}
Fig.~\ref{fig12} shows that the above expressions are indeed in agreement with the results obtained in the numerical simulations. 

Notice that the stability of the SVP is directly related to the maximal value of the interaction force, $F_\text{sv}^\text{max}$. Therefore, we expect the threshold values to be enhanced for: i) smaller penetration depth $\lambda$ of the superconducting film, which concentrates the magnetic flux in smaller regions, thus increasing the SVP interaction; ii) reduced thickness of the insulating layer, which increases the magnetic field of the vortex acting on the FM plane; iii) stronger DMI in the FM film, which enlarges the core of the skyrmion, thus aligns the magnetization of the core with the stray field of the vortex, thereby increasing the SVP interaction.

\section{Conclusion}\label{Sec.IV}

Precisely controlled dynamics of magnetic skyrmions in chiral ferromagnets has become of great relevance for cutting-edge memory devices and information technology applications. In this work, we described the resultant behavior of magnetic skyrmions when coupled to superconducting vortices in ferromagnet-superconductor hybrid systems. We have demonstrated that such a hybrid system enables multiple possibilities for manipulating the skyrmion-vortex pair, that are not possible for either constituent. We analyzed the dependence of the skyrmion-vortex coupled motion on the effective material viscosities, the exerted Lorentz-like force on vortices, and magnetic torques acting on a skyrmion, and determined the threshold values of external drives for which the skyrmion-vortex pair remains bound. Futhermore, we have calculated the Hall-angle of the skyrmion-vortex pair with respect to currents applied into either, or both superconducting and ferromagnetic films, and have thereby demonstrated the unprecedented tunability of the direction of motion for skyrmions in this hybrid system. Bearing in mind the plethora of known manners for manipulating fluxonics in superconductors by nanostructuring \cite{moshchalkov2010nanoscience}, and possibilities for similar manipulations of skyrmions \cite{menezes2019deflection,stosic2017pinning,reichhardt2015quantized,fernandes2018universality,reichhardt2015shapiro,reichhardt2015magnus}, our work opens a research direction of hybridized dynamics in SC-FM systems that holds promise to reveal rich fundamental phases and applicable effects. 

\section*{Acknowledgements}
This work was supported by the Research Foundation - Flanders (FWO-Vlaanderen) and  Brazilian Agencies  Funda\c{c}\~{a}o de Amparo a Ci\^{e}ncia e Tecnologia do Estado de Pernambuco (FACEPE, under the grant No. APQ-0198-1.05/14), Coordena\c{c}\~{a}o de Aperfei\c{c}oamento de Pessoal de N\'{\i}vel Superior (CAPES) and Conselho Nacional de Desenvolvimento Cient\'{\i}fico e Tecnológico (CNPq).

\appendix

\section{Skyrmion-vortex interaction for superconducting films of arbitrary thicknesses}\label{AppB}

\begin{figure}[b]
\centering
\includegraphics[width=\linewidth]{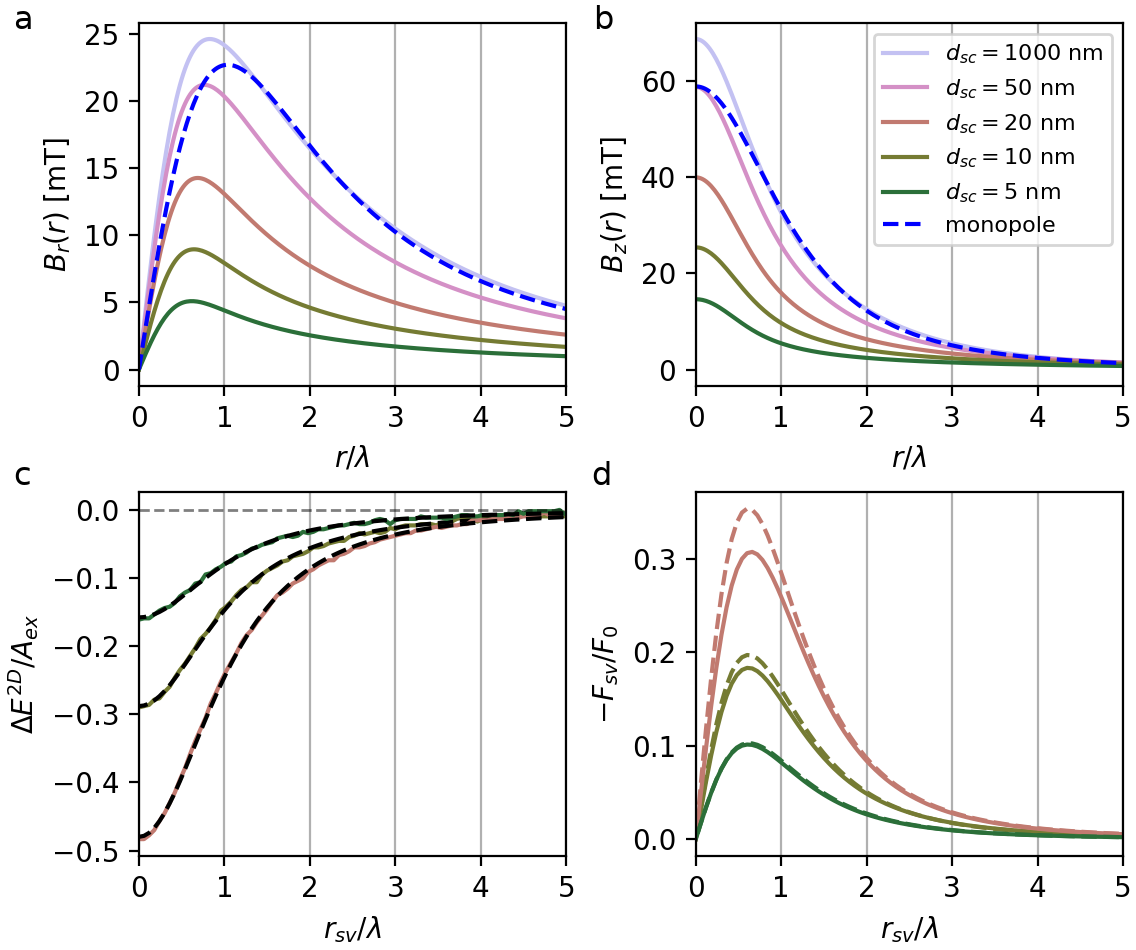}
\caption{(a,b) Stray magnetic field of the vortex for different thickness of the superconducting film, calculated in the plane of the FM film within the considered SC-FM hybrid. (c) Skyrmion-vortex interaction energy calculated in the micromagnetic simulations as a function of the distance between the skyrmion and the vortex cores, for $d_\text{SC}=5$, $10$ and $20$~nm. Here the energy curves were fitted by $E=a/(r_{\text{sv}}^2+b\lambda^2)^c$, with $a$, $b$, $c$ the fitting parameters (yielding black dashed lines). (d)  Corresponding interaction force calculated by the derivative of the fitted curves in (c), the dashed lines denote the pure-Zeeman component of the interaction force. In all calculations we take $\lambda=50$~nm, $d_\text{I}=10$~nm and $D=0.8D_c$. }
\label{fig13}
\end{figure}

In order to calculate the stray field of the vortex in a superconducting film of an arbitrary thickness $d_{sc}$, we integrate Eqs.~(\ref{eq0a}) and (\ref{eq0b}) numerically. Figs.~\ref{fig13} (a) and (b) show the obtained stray fields for different values of $d_\text{sc}$, with $\lambda=50$~nm and $d_\text{I}=10$~nm fixed, where we consider a finite vortex core by inserting the cutoff factor $\exp(-\xi^2k^2)$, with $\xi=10$~nm in Eqs.~(\ref{eq0a}) and (\ref{eq0b}). Figs.~\ref{fig13} (c) and (d) show the skyrmion-vortex interaction energy and interaction force, respectively, calculated as in Sec. \ref{S.III.C} of the main text, for $D=0.8D_c$. The dashed lines in Fig.~\ref{fig13} (d) show the pure-Zeeman component of the interaction force. Notice that even though for the considered parameters the skyrmion size is weakly affected by the presence of the vortex field, small changes in the skymion shape can still result in a non-negligible contribution of the non-Zeeman energy terms to the total skyrmion-vortex interaction. 

\section{Calculation of the dissipative tensor}

The dissipative tensor can be calculated by considering a single magnetic skyrmion with its center located at the origin $r=0$. The components of the dissipative tensor are defined as
\begin{equation}
\mathcal{D}_{ij}=\frac{d\Msat}{\gamma}\int d^2r\partial_i\textbf{m}\cdot\partial_j\textbf{m}.
\label{Eq.D0}
\end{equation}
The azimuthal symmetry of the spin configuration leads to $\mathcal{D}_{xx}=\mathcal{D}_{yy}=\mathcal{D}$ and $\mathcal{D}_{xy}=\mathcal{D}_{yx}=0$, and reduces the problem to a 1D integral
\begin{equation}
    \mathcal{D}=\frac{d\Msat}{\gamma}\pi\int_{0}^{\infty}rdr\left[\left(\frac{d\theta(r)}{dr}\right)^2 + \frac{\sin^2\theta(r)}{r^2} \right],
    \label{Eq.D2}
\end{equation}
where we used $\textbf{m}(\textbf{r})=\sin[\theta(r)]\hat{r}+\cos[\theta(r)]\hat{z}$ in Eq.~(\ref{Eq.D0}) for the case of a N\'eel skyrmion. Here $r=\sqrt{x^2+y^2}$ is the distance from the skyrmion core. Eq.~(\ref{Eq.D2}) can be discretized in the simulation as follows   
\begin{equation}
    \mathcal{D}=\frac{d\Msat}{\gamma}\pi\sum_{i=1}^{N}\left[\left(\frac{\theta(i+1)-\theta(i-1)}{2} \right)^2 + \frac{\sin^2\theta(i)}{i^2} \right],
    \label{Eq.D1}
\end{equation}
where $r=ia$, with $a$ the lattice separation. $N$ is such that $\xi_{sk}\ll Na$, with $\xi_{sk}$ the skyrmion radius.

For the results presented in Sec.~\ref{Sec.IIIb} we have calculated $\mathcal{D}\approx 2\times10^{-16}$~N/ms$^{-1}$, for the skyrmion at rest in the absence of applied fields, with $D=0.8D_c$ and the remaining FM parameters as given in Sec.~\ref{Sec.IIb}.

\bibliography{references}

\end{document}